# The Roads One Must Walk Down: Commute and Depression for Beijing's Residents

Xize Wang[a], Tao Liu[b, *]

**Abstract**

As a vital aspect of individual's quality of life, mental health has been included as an important component of the U.N. Sustainable Development Goals. This study focuses on a specific aspect of mental health: depression, and examines its relationship with commute patterns. Using survey data from 1,528 residents in Beijing, China, we find that every 10 additional minutes of commute time is associated with 1.1% higher likelihood of depression. We test for the mechanisms of the commute-depression link and find that commute is associated with depression as a direct stressor rather than triggering higher work stress. When decomposing commute time into mode-specific time, we found that time on mopeds/motorcycles has the strongest association with depression. Moreover, the commute-depression associations are stronger for older workers and blue-collar workers. Hence, policies that could reduce commute time, encourage work from home, improve job-housing balance or increase motorcyclists' safety would help promote mental health.

a. Department of Real Estate, National University of Singapore, Singapore. Email: wangxize316@gmail.com. OCRID: 0000-0002-4861-6002
b. College of Urban and Environmental Sciences and Center for Urban Future Research, Peking University. Email: wangxize316@gmail.com. OCRID: 0000-0002-3568-4882
* Corresponding author.

# 1 Introduction

Mental health is an important part of an individual's well-being and has been included as a component of the United Nations Sustainable Development Goals (Dickerson, Hole, and Munford 2014). In recent years, mental health issues have become an increasing social concern. For instance, in the United States, one in every five adults has experienced mental health issues at least once a year (Substance Abuse and Mental Health Services Administration 2012). In China, the latest nationwide survey to date, conducted in 2013-2015, reported a 6.8% lifetime prevalence rate of depressive disorders (Huang et al., 2019). Among China's urban residents, the prevalence of mental health issues is generally higher in big cities than small cities (Yang et al., 2018). The emergence of mental health issues has created a high financial and social burden in society. Studies on China show that the total direct medical cost of mental health issues in 2013 accounted for 15% of the total health expenditure, and 1.1% of China's GDP (Xu et al. 2016). Considering the possibility that mental health issues in China are often under-diagnosed (especially among women), and the fact that mental health issues also incur indirect costs such as productivity loss, the real costs would have been even higher (Gupta et al. 2015; Kessler et al. 2007).

Efforts to improve people's mental health will not only benefit the individual's quality of life but will also help to promote economic and social sustainability in cities. In addition to medical and psychiatric factors, environmental factors can also impact people's mental health (Marmot 2005). The urban physical environment, as a specific case of environmental factors, can influence people's mental health by constantly exposing people to stress (Caspi et al. 2003, Pearlin et al. 1981). In other words, planners and policy makers are able to promote urban residents' mental health conditions by improving the urban physical environment and creating suitable urban policies.

In the literature, mental health can either be measured by general mental well-being indices, or be examined in specific aspects including depression, stress, or others. This study focuses on a specific mental health condition: depression, and examines its relationship with a specific urban environment: commute durations. Among various mental health indicators, depression is frequently covered in psychotherapy practices and hence has a higher relevance to the healthcare system (Bradley, Bagnell, and Brannen 2010; Chen and Mui 2014); among different trip types, commute normally makes up a large share of people's total trip distance and trip durations (Boarnet, 2011; Liu, Ettema and Helbich, 2022). Existing studies have found that longer commute time is associated with worse mental health conditions (e.g., Feng and Boyle 2014; Wang et al. 2019). However, few studies have examined the role stress plays in such dynamics, and specifically, whether the association occurs through an increase in work stress. Regarding mode-specific effects, most studies focus on driving and active modes, and the effects of time spent walking and cycling on mental well-being are mixed (Humphreys, Goodman, and Ogilvie 2013; Martin, Goryakin, and Suhrcke 2014). Studies examining the effects of time spent on other modes (such as transit and mopeds), especially those systematically disentangle the overall commute time into mode-specific time to examine their effects, are needed. From a broader perspective, there is a growing body of literature examining the negative consequences of long commute time, such as life satisfaction and job productivity (Ma and Ye, 2019; Zhu et al. 2019). Studies on depression measures, which has relatively higher clinical relevance, are still relatively few.

To address these gaps, this study uses Beijing, China as a case study and examines the associations between door-to-door commute time and depressive symptoms for 1,582 survey respondents in urban and suburban neighborhoods. We also follow the "impedance theory" and the "stress process theory" and test whether commute time associates with depression by serving as a direct stressor, an indirect stressor that triggers higher work stress, or a moderator of the work stress-depression nexus. In addition, we decompose the overall commute time into mode-specific time to examine mode-specific effect sizes, and examine the mental health effect of multimodal commuting. Finally, we explore the heterogeneous effect levels for different age and occupation groups.

The remainder of the paper is organized as follows. The next section reviews the relevant literature. The third section describes the study area, the data, and the methodology. The fourth section presents the findings of the regression models. The fifth section discusses the findings and their implications for planners. The sixth section concludes the paper.

## 2 Literature review

The relationship between commute and mental health is supported both theoretically and empirically by the prior literature. This section reviews the relevant theoretical discussions and empirical findings that connect commute with different measurements of mental health: the more "clinically relevant" depression measures and other "less clinically relevant" ones.

From a theoretical perspective, the link between commute time and mental health originates in the "impedance theory" (Novaco et al. 1979; Novaco et al. 1990) and the "stress process theory" (Pearlin et al. 1981; Pearlin and Bierman 2013). The "impedance theory" aims to explain the psychophysiological link between commute and mental health. Specifically, this theory hypothesizes commute trips as psychological "impedance", which is the difficulty for a commuter in moving from origin to destination. Thus, commute can be regarded as an effort-taking process to overcome such "impedance" that creates stress (Novaco et al. 1979, Novaco et al. 1990). When people experience stress, their levels of cortisol increase, and those who constantly have high cortisol levels are more likely to have depression and other mental health problems (Adli 2017; Pearlin et al. 1981).

The "stress process theory" aims to explore the potential psychological mechanisms connecting commute and mental health, with a focus on the potential roles stress may play. Long commute time, according to the above-mentioned "impedance theory", can serve as an independent stressor to increase individuals' cortisol level, and hence increase the probability of depression (Novaco et al. 1979; Pearlin et al. 1981). Besides being such a "direct stressor", commute may also serve as an "indirect stressor" to trigger a "secondary stressor" that can lead to depression (Pearlin and Bierman 2013). For instance, long commute time may increase individuals' mental fatigue, anxiety and feeling of time pressure at the beginning of a workday, and hence increase their work stress level (Pearlin and Bierman 2013; Wener, Evans and Boatley 2015). Subsequently, the increased work stress level is further associated with higher likelihoods of depression (Ganster and Rosen 2013). Such commute-work stress and work stress-depression links are supported by recent empirical evidence (Amponsah-Tawiah et al. 2016; Ganster and Rosen 2013; Wener, Evans and Boatley 2005; Wang et al. 2021). Finally, the "stress process theory" also proposes a concept of "psychological resources", implying that certain non-psychophysiological factors can make a person more mentally robust towards stressors (Pearlin and Bierman 2013). For instance, certain urban environment factors such as shorter commute time can make an individual have higher ability to heal from work stress-induced cognitive fatigues; therefore, this individual may have relatively weaker work stress-depression associations (Evans, 2003).

Following these two theories, Figure 1 illustrates the three above-mentioned possible mechanisms whereby journey to work may be connected to depression: (a) the "direct stressor" hypothesis, (b) the "indirect stressor" hypothesis, and (c) the "moderator" hypothesis. Additionally, the relationship between commute and depression may differ among different sub-populations. For instance, this relationship may differ by age, as the level of mental fatigue after carrying out a task generally increases with age (Arnau et al. 2017; Pearlin and Bierman 2013). This relationship may differ by generations, as people in different birth cohorts may have different travel and lifestyle preferences (Garikapati et al., 2016; Wang and Wang 2021). This relationship may also differ by occupations, especially between the white-collar and blue-collar jobs (Dedele et al. 2019). This is because different occupations generally have different job quality, flexibility and self-control levels, which can further lead to mental health disparities (Wang et al. 2021). Note that although the conceptual models in Figure 1 focus on work stress due to its kinship with commute, the stress can also be non-work-related (Novaco et al. 1990; Turner et al. 1995). This study will test for both work stress and life stress, whereby the latter is discussed in Section 4.5 *"Robustness Checks"*.

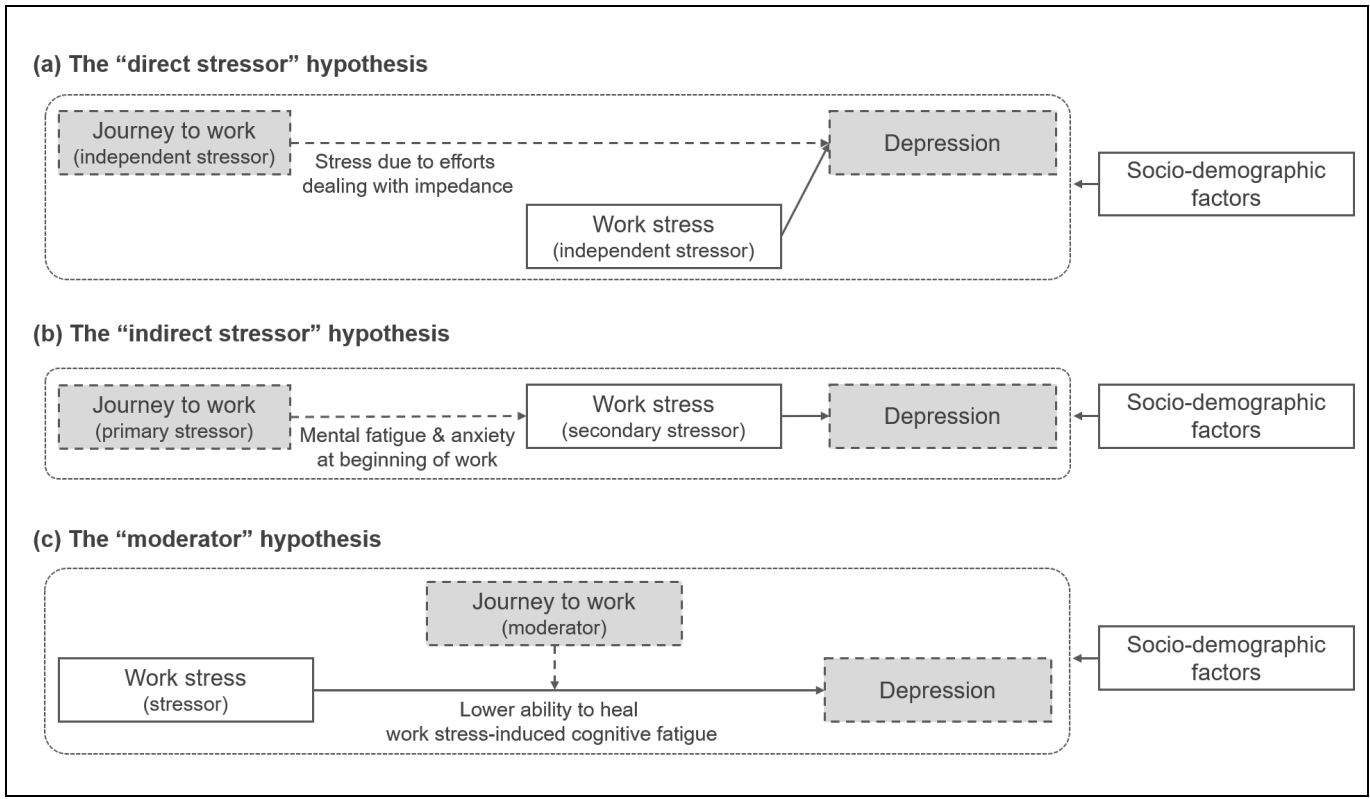

**Figure 1 - Conceptual models on commute, work stress, and depression**

Besides the theoretical discussions, recent empirical evidence has also documented the associations between commute time and mental health (Liu, Ettema and Helbich, 2022). For instance, using data from 5,216 participants in the British Household Panel Survey, Feng and Boyle (2014) find that long journeys to work are associated with a higher risk of poor mental health, measured by GHQ-12 indicators. Using survey data covering 5,438 participants across eleven Latin American cities, Wang et al. (2019) find that every 10 additional minutes spent commuting is associated with a 0.5% higher probability of having depressive symptoms as measured by the 10-item Center for Epidemiologic Studies Depression (CESD-10) scale. Besides GHQ-12 and CESD-10, recent studies also measure mental health using indicators such as MCS-8, MHI-5, PHQ-2 or SF-36 (Liu Ettema and Helbich, 2022). Studies in Australia, Canada and China also find statistically significant associations between longer commute time and worse mental health status (Hilbrecht et al. 2014; Milner et al. 2017; Xiao et al. 2020).

In addition to general mental well-being measures, existing studies also focus on stress. For instance, a study in Montreal, Canada surveyed 3,794 respondents in a university community and found that longer commuting time was associated with higher self-reported stress (Legrain, Eluru, and El-Geneidy 2015). Evans and Wener (2006) surveyed 208 train commuters in New York City and found that those with longer commute times had higher stress levels measured by levels of self-reported stress and salivary cortisol. Empirical evidence in Austria, Germany, Sweden, and United States also documents that individuals taking longer time to work has higher stress level during commute (Gottholmseder et al. 2009; Zhu and Fan 2018) and report higher levels of everyday stress (Hansson et al. 2011; Ruger et al. 2017).

For mode-specific effects, recent empirical evidence confirms that time spent in motorized modes is significantly associated with mental health status. For instance, studies show that longer commute time in cars or transit is associated with worse mental health conditions (Feng and Boyle 2014) and higher stress levels (Evans 2003, Evans and Wener 2006, Legrain, Eluru, and El-Geneidy 2015). In contrast, evidence regarding active modes is mixed. There are studies showing that active commute time is not significantly associated with general mental health, depression, or travel satisfaction (Humphreys, Goodman, and Ogilvie 2013, Kuwahara et al. 2015, Mao, Ettema, and Dijst 2016), and there is also evidence that those having longer walking time to work have higher stress levels (Legrain, Eluru, and El-Geneidy 2015). When comparing mental health across modes, non-motorized commuters have better mental health and lower stress than motorized commuters (Ma et al., 2021; Martin, Goryakin, and Suhrcke 2014), and transit commuters have lower stress than drivers (Wener and Evans 2011).

From a broader perspective, there is an emerging literature exploring the relationship between commute and well-being (Chatterjee et al. 2020; De Vos et al., 2013; Liu, Ettema and Helbich, 2022). Generally, longer commute time is associated with lower levels of well-being, including but not limited to mental health. Such well-being measures include life satisfactions (Fordham, van Lierop and El-Geneidy, 2018; Zhu et al. 2019), satisfactions with commute (De Vos, Ettema and Witlox, 2019; Gerber et al., 2020); affect (Morris and Guerra, 2015; Morris and Zhou 2018), happiness (Olsson et al., 2013; Yin et al. 2019), job productivity (Ma and Ye 2019), job satisfactions (Sun et al. 2021), and sleep time (Voulgaris, Smart, and Taylor 2019). In addition, active commuting is associated with higher levels of well-being (Chng et al. 2016; St. Lous et al., 2014), indicating the importance of designing walkable neighborhoods (Pfeiffer et al. 2020).

In summary, although theoretical and empirical studies regarding commute time and mental health are emerging, significant questions remain. First, although empirical evidence supports the theory that longer commute time is associated with higher stress and worse mental health, most studies only examined either stress or mental health alone. Studies examining commute time, stress, and mental health together can contribute to the literature by testing the mechanisms among them. Second, most studies examining the effects of mode-specific time only focused on one or two specific modes, such as driving or cycling. Studies that disentangle the overall commute time into mode-specific time and systematically examine their effects are still rare. Third, in the broad literature on commute time and subjective well-being, studies focusing on the clinically-relevant depressive symptoms are still relatively rare.

# 3 Data and methods

## 3.1 Survey data and study sample

The dataset for this study comes from a survey conducted in Beijing, China between November 2018 and April 2019. The survey covers seven of Beijing's 16 districts (Figure 2a): Xicheng in the old city (i.e., within the historical City Wall), Chaoyang, Fengtai, and Haidian within the Central City defined in the 2016 Beijing City Master Plan, and Fangshan, Tongzhou, and Changping outside the Central City. The survey follows a multistep stratified sampling probability proportional to size (PPS) sampling scheme. In each district, 2-5 subdistricts were randomly selected. In each subdistrict, 16-48 communities were randomly selected, and in each community, 10-20 households were randomly selected for the survey. There were 36 subdistricts, 168 communities, and 4,061 households that the survey covered. In each household, one main respondent was selected using the following criteria: (a) 18-59 years old, (b) has formal Beijing residency (i.e., *hukou*), and (c) lives in Beijing. While all household members were surveyed for basic demographic information, the main respondent was also surveyed with questions on commute patterns, mental health, work stress, home ownership, and income. There were 4,061 main respondents to this survey, and their sociodemographic profiles were comparable with those from the National 1% Population Survey 2015, the most recent national-level population survey available (National Bureau of Statistics of China 2016)[1].

---

[1] For the 4,061 survey respondents, the average age is 39.5 and the share of females is 48.6%; for the corresponding sample in the National 1% Population Survey, the average age is 39.9 and the share of females is 50.0%.

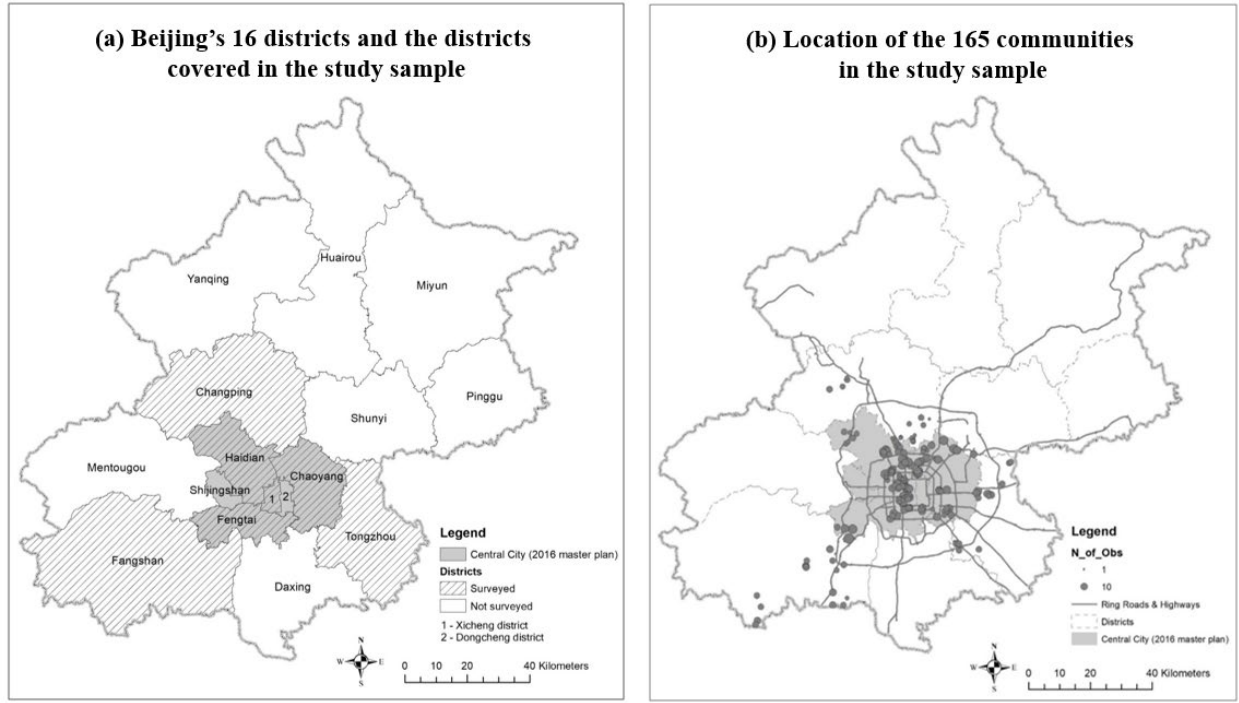

**Figure 2 – Districts and communities covered in the study sample**
Note: for (b), the size of points refers to the number of individuals in the community

The final study sample includes individuals who have full information on depression, work stress, commute patterns and control variables. The final study sample has 1,582 main survey respondents. 798 respondents were excluded due to lack of information on depression and stress, 1,251 respondents were excluded due to lack of complete information on commute variables, including missing total commute time, traffic delay time or mode-specific time, and 430 respondents were excluded due to lack of full sociodemographic control variables. The share of those screening positive for depression is higher for those in the study sample than for those outside of it (6.1% vs. 4.4%), and those in the study sample have slightly longer commute times than those outside it (39 minutes vs. 32 minutes). The final study sample spans over 165 of the 168 communities in the survey (Figure 2b).

*3.2   Depression, work stress, and commute variables*

The outcome variable, depressive symptoms, is measured by the 10-item Center for Epidemiologic Studies depression scale (CESD-10) (Andersen et al. 1994). This scale is frequently used in the healthcare sector to screen for depressive symptoms, and have been proven to be valid and reliable across various geographic and cultural contexts (Bradley, Bagnell, and Brannen 2010; Chen and Mui 2014). The CESD-10 scale includes a list of ten sentiments[2], and each individual is surveyed on how often in the past week that she/he had each of the ten sentiments (e.g., 1-2 days, 3-4 days, etc.). The responses are then converted to a score of 0-30, with higher values indicating a higher likelihood of depressive symptoms. A person with a CESD-10 score of 10 or higher is screened as positive for depression symptoms (Andersen et al. 1994). Hence, we use the scale to create a binary variable that equals one if the individual screens positively for depression and zero otherwise. All respondents with missing responses on the ten items were excluded.

Work stress is a binary variable transformed from the survey question on recent work stress levels. The original question has four categories: no stress, little stress, large stress and very large stress. For ease of estimating moderation effects, we re-categorized the scale into a binary variable with zero equal to no or little stress and one equal to "large" or "very large" stress. Hence, the binary variable measures whether the individual was recently experiencing significant work stress.

The key exposure variable is a self-reported door-to-door commute time during a normal day. We define the commute as the journey from home to the place for the main activity, including work, school, or other. In addition to the overall commute time, we further disentangle the total time into mode-specific time for multimode commuters. The survey asked the respondents to breakdown to overall, door-to-door commute time into eleven modes[3]. For instance, if a person takes 10 minutes to get from home to the subway station, 20 minutes on the subway, and another 5 minutes from the subway station to the workplace, this person would report 20 minutes in rail transit and 15 minutes walking. We group these eleven variables into five mode-specific commute time variables: time spent in transit (rail transit and bus), private auto (private auto only), moped/motorcycle (motorcycle/mopeds only), nonmotorized (walking, cycling using own bicycle, cycling using bike shares), and other (intercity rail, taxi, ride-hailing alone, ride-hailing carpool). For individuals not using any of these five mode groups, the time for that specific mode is recorded as zero. To examine the effects of multimodal commuting, we create a dummy variable indicating whether an individual has used at least two of the following modes: cycling, mopeds/motorcycles, intercity rails, rail transit, buses, private automobiles and taxi/ride-hailing. We exclude walking in the calculation, since normally walking plus another mode is not regarded as multimodal. Finally, to examine the effects of traffic delays, we create another set of commute variables including a hypothetical "free-flow" door-to-door commute time assuming no traffic and a traffic delay variable, which is the difference between total and free-flow commute time variables.

---

[2] The ten sentiments are: (1) I was bothered by things that usually don't bother me; (2) I had trouble keeping my mind on what I was doing; (3) I felt depressed; (4) I felt that everything I did was an effort; (5) I felt hopeful about the future; (6) I felt fearful; (7) My sleep was restless; (8) I was happy; (9) I felt lonely; (10) I could not "get going".

---

[3] These 11 modes are: walking, cycling (own bicycle), cycling (shared bicycle), motorcycle/mopeds, intercity rails, rail transit (subway/light rail), buses/shuttles, taxis, private automobiles (driver or passenger), ride-hailing (travel alone), and ride-hailing (carpooling).

*3.3   Control variables*

The control variables include nine socioeconomic profile variables and a binary variable for residing in the central city. The socioeconomic variables include gender (female or male), age in 2018 (years), living with a spouse/partner (y/n), having children (y/n), household-level income (in 10,000 Chinese Yuan or CNY), home ownership in Beijing (y/n, an important wealth indicator in the Chinese context), living in condominiums (y/n, with "no" indicating living in a variety of nonmarket housing types), level of education (categorical variable of lower than high school, high school/some college, and bachelor's degree or higher), and occupation (categorical variable of government/public agencies, managerial positions, professionals such as engineers/financial consultants, agriculture, manufacturing/construction, service and others).

The central city variable indicates whether the respondent lived in the six districts defined by the most recent 2016 Master Plan: Xicheng, Dongcheng, Haidian, Chaoyang, Fengtai, and Shijingshan (see Figure 2). This variable controls for the unobserved characteristics within and beyond the central city area. The spatial characteristics of the population and employment spatial distribution for the six central city districts were quite similar, and the largest difference emerged for the central city vs. non-central city (i.e. urban vs. suburban) comparisons as opposed to the within-district comparisons (Liu et al. 2021; Shi and Cao 2020). Ideally, district fixed effects variables would have been a better option than a central city dummy variable since it could control for district-level unobserved characteristics. However, due to the unbalanced distribution of those with and without depressive symptoms (6% vs. 94%), models using district fixed effects would drop cases due to lack of variation, especially for subsample regressions. As a robustness check, we re-run the full-sample regressions using district fixed effects, and the key findings remain unchanged (see Section 4.5 *"Robustness Checks"* for details).

*3.4  Descriptive statistics*

Table 1 shows the descriptive statistics of the study sample, which includes 1,582 individuals across 165 out of the 168 surveyed communities (see Figure 1). Among these 1,582 individuals, 96 or 6.1% screened positive for depressive symptoms. For those 96 individuals with depression, 59 resided in the central city (4.9% of the total central city respondents), and 37 resided outside the central city (9.6% of the total non-central city respondents). To our knowledge, there has been no census or other government-led surveys covering mental health questions. Alternatively, China Family Panel Studies (CFPS), a nationally representative survey led by the Institute of Social Science Survey of Peking University, has a related question in the 2018 wave. The question covers eight of the ten CESD-10 questions. Based on these eight questions, we calculate the respective eight-item CESD scores for our study sample and compare them with the Beijing *hukou* holders sample in the CFPS 2018 wave. The average eight-item CESD score in our study sample was lower than that of the CFPS subsample (3.76 vs. 4.99)[4].

For commute time, the average overall commute time is 38.9 minutes, comparable to the average commute time for Beijing *hukou* holders (39 minutes) according to the most recent nationwide 1% population survey in 2015 (National Bureau of Statistics of China 2016). The average traffic delay time is 4.4 minutes. For mode-specific time, transit has the highest average time of 20 minutes, followed by private automobiles with an average of 8.3 minutes. Non-motorized mode is the most popular mode: two thirds of the study sample reported using this mode, followed by transit (44%) and private auto (22%). For multimodal commuting, 14% of the study sample reported using more than one mode during their journey to work. For socioeconomic variables, 43% of the respondents are female, and the average age of the respondents is 38.3. The average annual household income is 222,660 Chinese yuan (or 33,382 US dollars). For occupation, professionals (engineering, finance or medicine) make up the largest share at 25%, and those working in agriculture make up the smallest share of 3.5%. For geographical distribution, 76% of the respondents reside in the central city defined by the 2016 Beijing Master Plan.

---

[4] Note that the average age of our study sample (38.3) is lower than that of the CFPS subsample (41.4, N=246).

Table 1 – Descriptive statistics of the study sample (N=1582)

|  | Mean or % | SD | Min | Max |
|---|---|---|---|---|
| *Dependent variable* | | | | |
| Depression (y/n) | 0.061 | 0.239 | 0 | 1 |
| | | | | |
| *Stress* | | | | |
| Work stress (y/n) | 0.598 | 0.490 | 0 | 1 |
| | | | | |
| *Commute time* | | | | |
| Overall one-way commute time (in 10 min) | 3.886 | 2.428 | 0 | 15 |
| Mode-specific time: transit (in 10 min) | 1.918 | 2.539 | 0 | 13 |
| Mode-specific time: private auto (in 10 min) | 0.825 | 1.793 | 0 | 12 |
| Mode-specific time: moped/motorcycle (in 10 min) | 0.273 | 0.824 | 0 | 5 |
| Mode-specific time: non-motorized (in 10 min) | 0.822 | 0.800 | 0 | 5 |
| Mode-specific time: others (in 10 min) | 0.050 | 0.419 | 0 | 12 |
| Free-flow commute time (in 10 min) | 3.447 | 2.112 | 0 | 14 |
| Traffic delay time (in 10 min) | 0.439 | 0.623 | 0 | 5 |
| | | | | |
| *Commute modes* | | | | |
| Using transit (y/n) | 0.441 | 0.497 | 0 | 1 |
| Using private auto (y/n) | 0.219 | 0.414 | 0 | 1 |
| Using moped/motorcycle (y/n) | 0.124 | 0.330 | 0 | 1 |
| Using non-motorized modes (y/n) | 0.666 | 0.472 | 0 | 1 |
| Using others (y/n) | 0.034 | 0.180 | 0 | 1 |
| | | | | |
| *Multimodal commuting* | | | | |
| Using more than one mode during commute[a] (y/n) | 0.143 | 0.350 | 0 | 1 |
| | | | | |
| *Socio-demographic characteristics* | | | | |
| Female | 0.434 | 0.496 | 0 | 1 |
| Age in 2018 (years) | 38.32 | 9.981 | 18 | 59 |
| Living with spouse/partner (y/n) | 0.639 | 0.480 | 0 | 1 |
| Having children (y/n) | 0.612 | 0.487 | 0 | 1 |
| Household income in 2017 (in 10k) | 22.17 | 33.93 | 1 | 1000 |
| Home ownership in city (y/n) | 0.566 | 0.496 | 0 | 1 |
| Living in condominium (y/n) | 0.487 | 0.500 | 0 | 1 |
| Education (%) | | | | |
|   *Less than High School* | 9.1% | | | |
|   *High school/some college* | 49.2% | | | |
|   *College or higher* | 41.7% | | | |
| Occupation (%) | | | | |
|   *Government* | 23.6% | | | |
|   *Manager* | 11.0% | | | |
|   *Professional* | 25.2% | | | |
|   *Service* | 19.9% | | | |
|   *Agriculture* | 03.5% | | | |
|   *Manufacture/construction* | 9.8% | | | |
|   *Other* | 7.0% | | | |
| | | | | |
| *Residential locations* | | | | |
| Inside central city (2016 plan) (y/n) | 0.755 | 0.430 | 0 | 1 |
| District of residence (%) | | | | |
|   *Xicheng* | 16.9% | | | |
|   *Chaoyang* | 20.8% | | | |
|   *Fengtai* | 11.6% | | | |

| | |
|---|---|
| *Haidian* | 26.2% |
| *Fangshan* | 9.0% |
| *Tongzhou* | 10.0% |
| *Changping* | 5.4% |

Note: a. This dummy variable equals one if the individual uses at least two of the following modes in commute: cycling, moped/motorcycle, intercity rails, rail transit, buses, private automobiles and taxi/ride-hailing.

*3.5  Statistical modeling*

To test for the three hypothesized relaionships among commute time, work stress, and depression, we propose three sets of regression models described by the following equations:

$$Depress_i = f(\beta_0 + \beta_1\ CommuteTime_i + \beta_2\ WorkStress_i + \beta_3\ SocioDem_i + \beta_4\ CentralCity_i), \quad (1)$$

$$Depress_i = f(\beta_0 + \beta_1\ CommuteTime_i + \beta_2\ SocioDem_i + \beta_3\ CentralCity_i), \quad (2)$$

$$WorkStress_i = f(\beta_0 + \beta_1\ CommuteTime_i + \beta_2\ SocioDem_i + \beta_3\ CentralCity_i), \quad (3)$$

$$Depress_i = f(\beta_0 + \beta_1\ CommuteTime_i + \beta_2\ WorkStress_i + \beta_3\ CommuteTime_i \times WorkStress_i \\ + \beta_7\ SocioDem_i + \beta_8\ CentralCity_i), \quad (4)$$

where $Depress_i$ measures whether individual $i$ is screened positive for depression, $CommuteTime_i$ measures individual $i$'s door-to-door commute time from home to workplace, and $WorkStress_i$ measures the individual's work stress. $SocioDem_i$ includes a set of sociodemographic control variables that are theoretically associated with both commute time and work stress. $CentralCity_i$ is a dummy variable indicating whether individual $i$ lives in Beijing's central city area defined by the 2016 Beijing Master Plan.

To test for the hypothesis in Row (a) of Figure 1 ("direct stressor"), we run the model in Equation (1) to determine if the coefficient of $CommuteTime_i$ was statistically significant after controlling for work stress, sociodemographic profiles, and central city indicators. To test for the hypothesis in Row (b) of Figure 1 ("indirect stressor" or mediation), we run the model in Equation (2) without $WorkStress_i$ to determine whether the coefficients of $CommuteTime_i$ in Equations (1) and (2) substantially differ; we also run the model in Equation (3) to determine whether $WorkStress_i$ and $CommuteTime_i$ are significantly associated. If the coefficients in Equations (1) and (2) differ and the coefficient in Equation (3) is statistically significant, it indicates that commute is associated with depression through increasing or decreasing work stress. To test for the hypothesis in Row (c) of Figure 1 ("moderator"), we run the model in Equation (3) to determine if the coefficient of the interaction term $CommuteTime_i \times WorkStress_i$ is statistically significant. If the coefficient is significant, it indicates that the commute time-depression association was stronger for those with longer commute time. As a robustness check, we replace the work stress variable with the more broadly-defined life stress, and this treatment does not affect the main results (see Section 4.5 *"Robustness Checks"* for details).

To examine the effects of mode-specific time, traffic delay time and multimodal commuting, we propose the following three additional regression models:

$$Depress_i = f(\beta_0 + \beta_1\ TransitTime_i + \beta_2\ CarTime_i + \beta_3\ NonMotorizedTime_i + \beta_4\ MopedTime_i + \beta_5\ OtherTime_i + \beta_6\ WorkStress_i + \beta_7\ SocioDem_i + \beta_8\ CentralCity_i), \quad (5)$$

$$Depress_i = f(\beta_0 + \beta_1\ FlowTime_i + \beta_2\ DelayTime_i + \beta_3\ WorkStress_i + \beta_4\ SocioDem_i + \beta_5\ CentralCity_i), \quad (6)$$

$$Depress_i = f(\beta_0 + \beta_1\ MultiModal_i + \beta_2\ WorkStress_i + \beta_3\ SocioDem_i + \beta_4\ CentralCity_i), \quad (7)$$

where $TransitTime_i$, $CarTime_i$, $NonMotorizedTime_i$, $MopedTime_i$ and $OtherTime_i$ measure commute time spent in transit, private automobiles, non-motorized modes, mopeds/motorcycles and other modes, respectively; $MultiModal_i$ is a dummy variable on multimodal commuting; and $FlowTime_i$ and $DelayTime_i$ are the hypothetical "free-flow" door-to-door commute time assuming no traffic and the traffic delay time during commute, respectively.

All models are logit regressions as the dependent variable is binary. We cluster the standard errors at the district level so that the possible within-district correlations would not overestimate the statistical significance of key variables. We note that when interpreting the coefficients, the marginal effects only apply within the variables' observed ranges.

## 4 Results

### 4.1 *Commute time, work stress, and depression*

Table 2 shows that total commute time is associated with depression, but this association is not mediated through work stress. Additionally, the work stress-depression association is not moderated by commute time. In other words, among the three hypotheses posited in Figure 1, the "direct stressor" hypothesis is supported by the models. Specifically, according to Table 2, Column (2), when holding continuous variables at their means and categorical variables at their modes, every 10 additional minutes in commute is associated with a 1.1% higher likelihood of having depression, and those with high work stress are 3.7% more likely to have depression than those with no or low work stress. The coefficients of overall commute time in Columns (1) and (2) are minimally different, and the coefficient commute time in Column (3) is not statistically significant, indicating that work stress is not a mediator of the commute time-depression association. In other words, long commute time is associated with depression as a direct stressor rather than triggering a higher level of work stress as an indirect stressor. In addition, in Table 2, Column (4), the interaction term between commute time and work stress is not statistically significant, indicating that the work stress-depression association did not vary by different levels of commute time.

Table 2 – Commute time, work stress, and depression

| | (1) Depression Commute Time Only | (2) Depression Commute Time and Work Stress | (3) Work stress Commute Time Only | (4) Depression Time-Stress Interaction |
|---|---|---|---|---|
| Overall one-way commute time (in 10 mins) | 0.222*** | 0.236*** | -0.067 | 0.252** |
| | [0.077] | [0.083] | [0.043] | [0.100] |
| Work stress (y/n) | | 1.244*** | | 1.347*** |
| | | [0.258] | | [0.425] |
| Commute time – work stress interaction | | | | -0.021 |
| | | | | [0.088] |
| Female | 0.286 | 0.450** | -0.174 | 0.448** |
| | [0.238] | [0.181] | [0.169] | [0.177] |
| Age in 2018 (years) | 0.042*** | 0.059*** | -0.055*** | 0.059*** |
| | [0.014] | [0.015] | [0.011] | [0.014] |
| Living with spouse/partner | -0.322 | -0.414 | -0.124 | -0.410 |
| | [0.659] | [0.592] | [0.363] | [0.593] |
| Having children | -0.561 | -0.766 | 0.851*** | -0.775 |
| | [0.529] | [0.599] | [0.310] | [0.589] |
| Household income in 2017 (in 10k CNY) | 0.003*** | 0.003*** | -0.005 | 0.003*** |
| | [0.001] | [0.001] | [0.008] | [0.001] |
| Home ownership in Beijing | 0.515 | 0.661 | -0.290 | 0.660 |
| | [0.648] | [0.665] | [0.252] | [0.663] |
| Living in condominium | -0.840*** | -0.915*** | 0.046 | -0.917*** |
| | [0.227] | [0.263] | [0.156] | [0.260] |
| Education | | | | |
| *Less than High School* | (ref.) | (ref.) | (ref.) | (ref.) |
| *High school/some college* | 0.562* | 0.555** | 0.191 | 0.543** |
| | [0.294] | [0.264] | [0.215] | [0.255] |
| *College or higher* | -0.020 | -0.069 | 0.050 | -0.082 |
| | [0.463] | [0.485] | [0.278] | [0.497] |
| Occupation | | | | |
| *Government* | (ref.) | (ref.) | (ref.) | (ref.) |
| *Manager* | 0.730** | 0.806** | 0.281 | 0.800** |
| | [0.347] | [0.318] | [0.303] | [0.321] |
| *Professional* | 0.045 | 0.172 | -0.156 | 0.170 |
| | [0.365] | [0.387] | [0.174] | [0.385] |
| *Service* | 0.878*** | 0.934*** | 0.022 | 0.929*** |
| | [0.236] | [0.227] | [0.206] | [0.237] |
| *Agriculture* | 0.040 | 0.226 | -0.451 | 0.199 |
| | [1.073] | [1.094] | [0.326] | [0.992] |
| *Manufacture/construction* | 0.964** | 1.022** | 0.110 | 1.017** |
| | [0.421] | [0.414] | [0.252] | [0.414] |
| *Other* | 1.909*** | 2.391*** | -1.421** | 2.399*** |
| | [0.226] | [0.268] | [0.592] | [0.277] |
| Inside central city (2016 plan) | -0.366 | -0.259 | -0.822** | -0.260 |
| | [0.477] | [0.497] | [0.324] | [0.498] |
| Constant | -5.719*** | -7.399*** | 3.305*** | -7.476*** |
| | [1.033] | [1.224] | [0.739] | [1.198] |
| Pseudo-R2 | 0.14 | 0.17 | 0.07 | 0.17 |
| Number of observations | 1582 | 1582 | 1582 | 1582 |

Note: Logit regression with screening positive for depression as the dependent variable. "Inside central city" indicates respondents living in the central city area designated by the 2016 Beijing Master Plan (i.e., Xicheng,

Dongcheng, Haidian, Chaoyang, Fengtai and Shijingshan Districts). *, **, *** indicate significance at the 0.10, 0.05 and 0.01 levels, respectively. Standard errors (clustered at the district level) are in parentheses.

*4.2    Effects of mode-specific time, traffic delay time, and multimodal commuting*

Table 3 disentangles the overall commute time into mode-specific time in Column (2) and free-flow/traffic delay time in Column (3), with Column (1) showing the full model (Table 2, Column (2)) as a reference. Table 3 also include a regression model with a dummy variable on multimodal commuting in Column (4). Column (2) shows that time spent in transit, private automobiles, and mopeds/motorcycles are significantly associated with depression, whereas the time spent in non-motorized modes and time spent in other modes are not significantly associated with depression. Holding other factors at their means (continuous) or modes (categorical), every 10 additional minutes in transit, private auto, and mopeds travel is associated with 1.1%, 1.3%, and 1.9% higher likelihood of depression, respectively. Note that the marginal effect for time spent in the moped/motorcycle is the highest among the five mode-specific time variables, but only 12.4% of the study sample reported using a moped or motorcycle in their commute (see Table 1). Column (3) shows that free-flow time, rather than traffic delay time, is significantly associated with depressive symptoms. Holding all continuous variables at their means and categorical variables at their modes, every 10 additional minutes in free-flow time is associated with a 1.2% higher likelihood of having depression. Column (4) shows that there is no significant relationship between multimodal commuting and depression, although the coefficient of the multimodal commuting dummy variable is positive.

**Table 3 - Effects of mode-specific time and free-flow/delay time on depression**

| | (1) Depression Overall Time | (2) Depression Mode-Specific Time | (3) Depression Free-flow time vs. delay time | (4) Depression Multimodal Commuting |
|---|---|---|---|---|
| Overall one-way commute time (in 10 mins) | 0.236*** [0.083] | | | |
| Mode-specific time: transit (in 10mins) | | 0.250*** [0.082] | | |
| Mode-specific time: private auto (in 10 min) | | 0.290*** [0.095] | | |
| Mode-specific time: non-motorized (in 10 min) | | 0.284 [0.250] | | |
| Mode-specific time: moped/motorcycle (in 10 min) | | 0.431** [0.170] | | |
| Mode-specific time: others (in 10 min) | | 0.033 [0.097] | | |
| Using more than one mode during commute (y/n) | | | | 0.409 [0.372] |
| Free-flow commute time (in 10 mins) | | | 0.253*** [0.069] | |
| Traffic delay time (in 10 mins) | | | 0.142 [0.232] | |
| Work stress (y/n) | 1.244*** [0.258] | 1.280*** [0.280] | 1.234*** [0.269] | 1.129*** [0.229] |
| Female | 0.450** [0.181] | 0.455** [0.189] | 0.447** [0.182] | 0.393** [0.169] |
| Age in 2018 (years) | 0.059*** [0.015] | 0.058*** [0.014] | 0.059*** [0.015] | 0.050*** [0.017] |
| Living with spouse/partner | -0.414 [0.592] | -0.486 [0.582] | -0.420 [0.597] | -0.148 [0.638] |
| Having children | -0.766 [0.599] | -0.739 [0.601] | -0.739 [0.640] | -0.913 [0.610] |
| Household income in 2017 (in 10k CNY) | 0.003*** [0.001] | 0.003*** [0.001] | 0.003*** [0.001] | 0.003*** [0.001] |
| Own home in Beijing | 0.661 [0.665] | 0.718 [0.677] | 0.654 [0.664] | 0.540 [0.634] |
| Living in condominium | -0.915*** [0.263] | -0.910*** [0.251] | -0.932*** [0.236] | -0.950*** [0.317] |
| Education | | | | |
| *Less than High School* | (ref.) | (ref.) | (ref.) | |
| *High school/some college* | 0.555** [0.264] | 0.476* [0.284] | 0.560** [0.262] | 0.800*** [0.222] |
| *College or higher* | -0.069 [0.485] | -0.159 [0.488] | -0.065 [0.479] | 0.384 [0.427] |
| Occupation | | | | |
| *Government* | (ref.) | (ref.) | (ref.) | |
| *Manager* | 0.806** [0.318] | 0.779** [0.333] | 0.806** [0.316] | 1.006*** [0.267] |
| *Professional* | 0.172 [0.387] | 0.158 [0.378] | 0.164 [0.396] | 0.246 [0.420] |
| *Service* | 0.934*** [0.227] | 0.964*** [0.261] | 0.919*** [0.248] | 0.954*** [0.211] |

|  | (1) | (2) | (3) | (4) |
|---|---|---|---|---|
| *Agriculture* | 0.226 | 0.288 | 0.180 | 0.029 |
|  | [1.094] | [1.101] | [1.026] | [1.167] |
| *Manufacture/construction* | 1.022** | 1.019** | 1.012** | 1.100*** |
|  | [0.414] | [0.413] | [0.425] | [0.395] |
| *Other* | 2.391*** | 2.419*** | 2.376*** | 2.390*** |
|  | [0.268] | [0.243] | [0.290] | [0.255] |
| Inside central city (2016 plan) | -0.259 | -0.286 | -0.262 | -0.196 |
|  | [0.497] | [0.477] | [0.507] | [0.525] |
| Constant | -7.399*** | -7.492*** | -7.398*** | -6.374*** |
|  | [1.224] | [1.201] | [1.233] | [1.015] |
| Pseudo-R2 | 0.17 | 0.18 | 0.17 | 0.13 |
| Number of observations | 1582 | 1582 | 1582 | 1582 |

Note: Logit regression with screening positive for depression as the dependent variable. "Inside central city" indicates respondents living in the central city area designated by the 2016 Beijing Master Plan (i.e., Xicheng, Dongcheng, Haidian, Chaoyang, Fengtai and Shijingshan Districts). Dummy variable "using more than mode during commute" equals one if the individual uses at least two of the following modes in commute: cycling, moped/motorcycle, intercity rails, rail transit, buses, private automobiles and taxi/ride-hailing. *, **, *** indicate significance at the 0.10, 0.05 and 0.01 levels, respectively. Standard errors (clustered at the district level) are in parentheses.

*4.3   Age and job-specific effects*

The subsample regressions in Table 4 show that the commute-depression association is stronger for those with older age and those working in blue-collar jobs. As indicated in Columns (2) and (3), commute time is significantly associated with depression for both the "up to 40 years old" and "41 and older" subsamples. However, the marginal effects for the "41 and older" subsample are larger than those of the younger subsample. Holding other factors at their means or modes for the study sample, every 10 additional minutes of overall commute time was associated with a 5.2% higher likelihood of depression for the "41 and older" subsample, whereas that of the "up to 40" subsample was only 0.5%. Models in Columns (4) and (5) include subsamples of individuals working in "white-collar" and "blue-collar" occupations, respectively, where "white-collar" jobs include government, managerial, and professional jobs, and "blue-collar" jobs include agricultural, manufacturing/construction, and service jobs. Column (4) shows that for "blue-collar" workers, commute time is significantly associated with depression, and holding other factors at means or modes, 10 additional minutes of commute time was associated with a 2.3% higher likelihood of depression. On the other hand, the commute time-depression association for "white-collar" workers was only significant at the 10% level.

**Table 4 – Commute time and depression by age group and by job type**

| | (1) Depression All | (2) Depression 40 & under | (3) Depression 41 & over | (4) Depression White-collar | (5) Depression Blue-collar |
|---|---|---|---|---|---|
| Overall one-way commute time (in 10 mins) | 0.236*** | 0.160** | 0.345*** | 0.196* | 0.272** |
| | [0.083] | [0.075] | [0.119] | [0.113] | [0.106] |
| Work stress | 1.244*** | 1.198*** | 1.148*** | 1.759*** | 1.455*** |
| | [0.258] | [0.239] | [0.421] | [0.667] | [0.254] |
| Female | 0.450** | 0.430 | 0.295 | 0.076 | 0.925*** |
| | [0.181] | [0.292] | [0.292] | [0.466] | [0.351] |
| Age in 2018 (years) | 0.059*** | | | 0.043*** | 0.082* |
| | [0.015] | | | [0.015] | [0.042] |
| Living with spouse/partner | -0.414 | 0.711 | -1.504*** | 1.407** | -1.904*** |
| | [0.592] | [0.490] | [0.346] | [0.650] | [0.611] |
| Having children | -0.766 | -1.046* | -0.357 | -1.764*** | 0.322 |
| | [0.599] | [0.619] | [0.507] | [0.476] | [1.131] |
| Household income in 2017 (in 10k) | 0.003*** | 0.001 | 0.008 | 0.002 | 0.011 |
| | [0.001] | [0.001] | [0.010] | [0.002] | [0.008] |
| Own home in Beijing | 0.661 | 0.510 | 0.788 | 1.224* | 0.730 |
| | [0.665] | [0.830] | [0.669] | [0.687] | [0.879] |
| Living in condominium | -0.915*** | -1.016** | -0.908*** | -0.855* | -1.199** |
| | [0.263] | [0.504] | [0.187] | [0.451] | [0.583] |
| Education | | | | | |
| *Less than High School* | (ref.) | (ref.) | (ref.) | (ref.) | (ref.) |
| *High school/some college* | 0.555** | 1.004 | 0.190 | 0.284 | 0.316 |
| | [0.264] | [1.547] | [0.400] | [1.171] | [0.551] |
| *College or higher* | -0.069 | 0.698 | -1.140** | -0.482 | -0.034 |
| | [0.485] | [1.782] | [0.480] | [0.884] | [0.820] |
| Occupation | | | | | |
| *Government* | (ref.) | (ref.) | (ref.) | | |
| *Manager* | 0.806** | 1.509*** | 0.259 | | |
| | [0.318] | [0.450] | [0.721] | | |
| *Professional* | 0.172 | 0.368 | -0.034 | | |
| | [0.387] | [0.508] | [0.502] | | |
| *Service* | 0.934*** | 0.506 | 1.065** | | |
| | [0.227] | [0.478] | [0.464] | | |
| *Agriculture* | 0.226 | 1.562 | -0.608 | | |
| | [1.094] | [1.243] | [1.012] | | |
| *Manufacture/construction* | 1.022** | 1.629** | 0.344 | | |
| | [0.414] | [0.753] | [0.414] | | |
| *Other* | 2.391*** | 2.942*** | 2.111*** | | |
| | [0.268] | [0.405] | [0.508] | | |
| Inside central city (2016 plan) | -0.259 | -0.096 | -0.400 | 0.162 | 0.080 |
| | [0.497] | [0.652] | [0.412] | [0.637] | [0.605] |
| Constant | -7.399*** | -6.344** | -3.768*** | -7.616*** | -7.908*** |
| | [1.224] | [2.793] | [1.288] | [1.473] | [1.124] |
| Pseudo-R2 | 0.17 | 0.17 | 0.21 | 0.17 | 0.17 |
| Number of observations | 1582 | 930 | 652 | 945 | 526 |

Note: Logit regression with screening positive for depression as the dependent variable. "Inside central city" indicates respondents living in the central city area designated by the 2016 Beijing Master Plan (i.e., Xicheng, Dongcheng, Haidian, Chaoyang, Fengtai and Shijingshan Districts). "White collar jobs" include those working in public service, managerial, and professional positions; "blue collar jobs" include those working in the primary/agricultural, secondary/manufacturing, and tertiary/service industries. *, **, *** indicate significance at the 0.10, 0.05 and 0.01 levels, respectively. Standard errors (clustered at the district level) are in parentheses.

*4.4    Robustness checks*

To test for the robustness of our findings, we conducted a few additional sensitivity analyses. First, to test the robustness of the CESD-10 cutoff score (i.e., 10 or above for depression), we reran our main regression models (Table 2) by changing the cutoff score to 9 and 11 instead of 10. The sign and significance of the coefficients of the exposure variables remain largely unchanged. Second, we estimated negative binomial models with the raw CESD-10 score (ranging from 0 to 30) as the dependent variable, and the results were every similar to the logit models using the binary depression variable. Third, we expanded our sample by including individuals reporting nine items (i.e., missing one item) in the CESD-10 questionnaire. For these additional 14 individuals, we followed Andersen et al. (1994) and multiplied their raw CESD score by 1.11 to scale them from 0~27 to 0~30 and then applied the 10-score cutoff point. Regression models using this relaxed sample yield similar results. Fourth, we included 98 additional individuals who reported only overall commute time but did not report free-flow or mode-specific time and reran the main regression models, and the results remained unchanged. Fifth, to better control for the unobserved characteristics by different geographies, we replaced the central city binary variable with district-level fixed effects, and the modeling results remained the same. Sixth, we replaced the work stress variable with a more broadly defined life stress variable of a 0~15 stress score for living cost, housing, childbearing, work, and parent-support stress, and using this more broadly defined stress variable did not affect the main results. Finally, we estimated the variance inflation factor (VIF) for the regression models and did not find evidence of multicollinearity (for the main model, average VIF is 1.67, maximum VIF is 3.86 for the variable "having children", VIF for the exposure variable "commute time" is 1.13).

## 5    Discussion

Using Beijing as a case study, we find that longer commute time is associated with a higher risk of having depressive symptoms. Among the three possible mechanisms implied by the "impedance theory" and the "stress process theory", we find that efforts to overcome the "impedance" during commuting serves as a direct stressor rather than triggering a secondary stressor (e.g. higher work stress), which eventually contributes to depression; and we do not find evidence that commute time can work as a buffer to attenuate work stress-depression associations. We further disentangle the overall commute time and found that (a) time spent in transit, private automobile and moped/motorcycle are significantly associated with depression, and the marginal effect of time spent in a moped is 73% higher than that of transit and 46% higher than that of private cars. Time spent in non-motorized modes is not significantly associated with depression. (b) Free-flow time, as opposed to traffic delay time, is significantly associated with depressive symptoms. Yet we do not find a statistically significant association between multimodal commuting and depression. In addition, subsample regressions show that the commute time-depressive symptom association is stronger for older respondents and blue-collar workers.

The finding that longer commuting time is positively associated with depressive symptoms is in agreement with previous studies in different urban and cultural settings (e.g., (Feng and Boyle 2014; Wang et al. 2019)). Studies using general mental health measures (e.g. GHQ-12) covers various aspects of mental health such as social functioning, depression and confidence. In contrast, this study only focuses on the "depression aspect" of mental health, which has a higher relevance to the healthcare system. Our findings that commute time contributes to depression as a direct stressor, not by triggering higher work stress, or moderating the work stress-mental health relationship can extend the theoretical and empirical literature that connects travel behavior and stress (e.g., (Avila-Palencia et al. 2017, Evans and Wener 2006, Novaco et al. 1979)). Such findings could connect with prior studies that establish a significant relationship between longer commute time and higher levels of commuting stress (Evans and Wener 2006; Gottholmseder et al. 2009). Unfortunately, our survey does not include information on commuting stress levels. Hence, we are unable to further explore whether the commute time – depression association is through increased commuting stress, which should be addressed in future research.

We find that time spent in a moped (or motorcycle) has the highest marginal effects on depressive symptoms, followed by time in cars and time in transit. This finding implies that driving mopeds/motorcycles is a strong contributor to mental health issues. In Beijing (and most other Chinese cities), there are no dedicated motorcycle lanes; moped and motorcycle users can use both the automobile and bicycle lanes. Hence, a possible explanation of this finding is that driving mopeds takes relatively higher mental effort than other modes due to the mopeds' ability to cut through vehicles and bicycles on congested roads. Additionally, driving mopeds may generate higher levels of stress due to mopeds' relatively higher risk of traffic crashes. This situation may also explain the slightly higher marginal effect for time in a car than time in transit since driving takes more mental effort than taking trains and buses.

We do not find significant associations between time spent in non-motorized modes and depression. Such findings are consistent with previous studies, such as Humphreys, Goodman, and Ogilvie (2013). The insignificance of active travel time is likely to be a composite effect of two different forces. On the one hand, longer time spent in the journey to work is associated with a higher likelihood of depression, but on the other hand, longer time being physically active is associated with a lower likelihood of depression (e.g., (Hong et al. 2009)). Although the net effect is ambiguous, it is possible that the insignificance of this variable is an outcome of these two counteracting effects canceling out each other. The insignificance of walking/biking time implies that an active mode to work may promote, or at least will not harm, one's mental health.

This study finds that free-flow time, as opposed to traffic delay time, is significantly associated with depressive symptoms. This finding differs from one author's previous study on 11 Latin American cities, in which traffic delay time is significant but free-flow time is not (Wang et al. 2019). One possible explanation is that since Beijing has an established rail transit system with 23 lines, 404 stations, and 699 km of railways, most people taking buses or private vehicles will only do so if they do not expect many delays. Alternatively, the time waiting for transit (due to long waiting lines during peak hours) could be the counterpart of Latin America's traffic delay time and significantly contribute to depression. Unfortunately, we currently do not have data to explore this hypothesis and plan to examine this issue in follow-up studies.

We do not find a significant association between multimodal commuting and higher likelihood of depression, although the sign of the coefficient is positive. Theoretically, multimodal commuting, especially that involves intermodal transfers, should create a higher level of impedance and lead to higher likelihood of depression (Navaco et al. 1979; Navaco et al. 1990). A possible explanation is that the survey data only includes mode-specific time but does not include number of transfers. In other words, our study sample cannot differentiate between (a) those spending 30 minutes in one single bus with no transfers and (b) those taking three bus trips with 10 minutes each. Future studies should address this question by collecting appropriate data.

Our subsample models identify two subgroups that are more sensitive to commuting as a contributor to depression. The first subgroup is those in their 40s or older. One potential explanation is that as people age, they are more prone to fatigue related to commuting or other activities (Arnau et al. 2017). Another possible explanation is that people in different generations may have different attitudes towards travel and have different expectations of the ideal commute time and modes (Wang 2019). The second subgroup is blue-collar workers. One possible explanation is that blue-collar workers are more likely to take fatigue-prone modes. For instance, in our study sample, 18% of the blue-collar workers reported using mopeds or motorcycles in their commute, while the share for white-collar workers is only 10%. Another possible explanation is that blue collar workers have a more fixed work schedule than white collar workers, and as a result, their commute experience is less likely to be by choice than white collar workers (Dedele et al. 2019; Wang et al. 2021).

The findings of this study support and highlight the importance of urban and transportation planning in promoting the mental health of urban residents. Efforts to reduce commute time and to optimize transportation network performance will help promote the mental health of residents. For instance, a flexible work schedule and work at home for one day a week should reduce the average commute time per day and hence decrease the probability of depression. Policies and programs to reduce the use of private automobiles and motorcycles and to promote the use of non-motorized modes and transit would also be beneficial to people's mental health. Planners and policy makers should pay special attention to those using mopeds or motorcycles in their commute, as they seem to be the most vulnerable group for converting long commute time to mental health issues. In addition to traffic safety programs for moped or motorcycles users, policy makers and planners should consider providing dedicated lanes for the users.

In addition, policy makers should pay attention to blue collar workers and those with relatively higher age, as these workers are relatively more vulnerable to long commute times contributing to depression. Among the many other previously mentioned policy suggestions, programs to improve the job-housing balance for older blue-collar workers deserve attention. For instance, the Beijing Municipal Government has started a joint-ownership public housing program (*"Gong you chan quan zhu fang")*, which offers lower prices but has strong regulations on eligibility to purchase and permission to transfer. Such joint-ownership estates are normally located at the fringe of central city areas, have good access to the rail transit system and have amenities specifically fit for the lower middle class lifestyle. Hence, such a housing program would be especially helpful for older and blue-collar workers to live in a place that has reasonable transit access and, among other benefits, reduces the mental health risks of their journey to work.

This study has the following limitations, many of which should motivate future research. First, our study sample comes from a cross-sectional dataset, and the relationships found in the study can only be interpreted as associational rather than causal; further studies should consider collecting longitudinal datasets or utilizing natural experiments. Second, this study focuses on those with "formal residency" in Beijing, which means *hukou* holders, who constitute only approximately two-thirds of its total population (Beijing Municipal Bureau of Statistics 2020). It is possible that migrant workers will have different effects than *hukou* holders. Third, overall commute time, uncongested travel time, and mode-specific commute time variables are all self-reported, which might be different from the objective measures such as those from mobile phone signals. Fourth, although we rule out the possibility that commute time is associated with depression through increased work stress or life stress, we are not able to test for whether the commute time – depression association is through commute stress or other mediators due to data limitations. Future research should collect appropriate data and examine this issue. Finally, although the geographic distribution of those in and outside the study sample is comparable (see Figure 2), the individuals included in and excluded from the study sample may be different. Although controlling for socioeconomic variables can adjust for the bias created by these variables (King, Keohane, and Verba 1994), they may also differ in unobserved ways.

# 6 Conclusions

Using self-reported data from Beijing's residents, this study find that a longer time spent in the journey to work was associated with a higher probability of depression. Specifically, long commute time contributes to depression as a direct stressor rather than triggering higher levels of work stress. We then disentangle the total commute time into mode-specific time and find that time spent in mopeds/motorcycles has the largest marginal effect, while time spent in a non-motorized mode has no significant effects. Those who are relatively older and working in blue-collar jobs are more sensitive with respect to the journey-to-work-depression associations. For planners and policy makers, programs and policies that reduce commute time, promote active travel, encourage work at home, promote job-housing balance and increase the safety of mopeds will help improve the mental health of urban residents.

**References**

Adli, Mazda. 2017. "Stress and the city." *Warum Städte uns krank machen. Und warum sie trotzdem gut für uns sind. München: C. Bertelsmann*.

Amponsah-Tawiah, K., Annor, F., & Arthur, B. G. (2016). Linking commuting stress to job satisfaction and turnover intention: The mediating role of burnout. *Journal of Workplace Behavioral Health*, *31*(2), 104-123.

Andersen, EM, Judith A Malmgren, William B Carter, and Donald L Patrick. 1994. "Screening for depression in well older adults: Evaluation of a short form of the CES-D." *American journal of preventive medicine* 10 (2):77-84.

Arnau, S., Möckel, T., Rinkenauer, G., & Wascher, E. (2017). The interconnection of mental fatigue and aging: An EEG study. *International Journal of Psychophysiology*, *117*, 17-25.

Avila-Palencia, Ione, Audrey de Nazelle, Tom Cole-Hunter, David Donaire-Gonzalez, Michael Jerrett, Daniel A Rodriguez, and Mark J Nieuwenhuijsen. 2017. "The relationship between bicycle commuting and perceived stress: a cross-sectional study." *BMJ open* 7 (6):e013542.

Beijing Municipal Bureau of Statistics. 2020. Beijing Economic and Social Statistical Communique 2019.

Boarnet, M. G. (2011). A broader context for land use and travel behavior, and a research agenda. *Journal of the American Planning Association*, *77*(3), 197-213.

Bradley, Kristina L., Alexa L. Bagnell, and Cyndi L. Brannen. 2010. "Factorial Validity of the Center for Epidemiological Studies Depression 10 in Adolescents." *Issues in Mental Health Nursing* 31 (6):408-412.

Caspi, Avshalom, Karen Sugden, Terrie E Moffitt, Alan Taylor, Ian W Craig, HonaLee Harrington, Joseph McClay, Jonathan Mill, Judy Martin, and Antony Braithwaite. 2003. "Influence of life stress on depression: moderation by a polymorphism in the 5-HTT gene." *Science* 301 (5631):386-389.

Chatterjee, Kiron, Samuel Chng, Ben Clark, Adrian Davis, Jonas De Vos, Dick Ettema, Susan Handy, Adam Martin, and Louise Reardon. 2020. "Commuting and wellbeing: a critical overview of the literature with implications for policy and future research." *Transport reviews* 40 (1):5-34.

Chen, Huajuan, and Ada C Mui. 2014. "Factorial validity of the Center for Epidemiologic Studies Depression Scale short form in older population in China." *International psychogeriatrics* 26 (1):49.

Chng, S., M. White, C. Abraham, and S. Skippon. 2016. "Commuting and wellbeing in London: The roles of commute mode and local public transport connectivity." *Prev Med* 88:182-8.

Dedele, A., Miskinyte, A., Andrusaityte, S., & Bartkute, Z. (2019). Perceived stress among different occupational groups and the interaction with sedentary behaviour. *International journal of environmental research and public health*, *16*(23), 4595.

De Vos, J., Ettema, D., & Witlox, F. (2019). Effects of changing travel patterns on travel satisfaction: A focus on recently relocated residents. *Travel Behaviour and Society*, *16*, 42-49.

De Vos, J., Schwanen, T., Van Acker, V., & Witlox, F. (2013). Travel and subjective well-being: A focus on findings, methods and future research needs. *Transport Reviews*, *33*(4), 421-442.

Dickerson, A., A. R. Hole, and L. A. Munford. 2014. "The relationship between well-being and commuting revisited: Does the choice of methodology matter?" *Regional Science and Urban Economics* 49:321-329.

Evans, Gary W. 2003. "The built environment and mental health." *Journal of Urban Health* 80 (4):536-555.

Evans, Gary W., and Richard E Wener. 2006. "Rail commuting duration and passenger stress." *Health Psychology* 25 (3):408.

Feng, Zhiqiang, and Paul Boyle. 2014. "Do Long Journeys to Work Have Adverse Effects on Mental Health?" *Environment and Behavior* 46 (5):609-625.

Fordham, L., Lierop, D. V., & El-Geneidy, A. (2018). Examining the relationship between commuting and it's impact on overall life satisfaction. In *Quality of life and daily travel* (pp. 157-181). Springer, Cham.

Ganster, D. C., & Rosen, C. C. (2013). Work stress and employee health: A multidisciplinary review. *Journal of management*, *39*(5), 1085-1122.

Garikapati, V. M., Pendyala, R. M., Morris, E. A., Mokhtarian, P. L., & McDonald, N. (2016). Activity patterns, time use, and travel of millennials: a generation in transition?. *Transport Reviews*, *36*(5), 558-584.

Gerber, P., El-Geneidy, A., Manaugh, K., & Lord, S. (2020). From workplace attachment to commuter satisfaction before and after a workplace relocation. *Transportation Research Part F: Traffic Psychology and Behaviour*, *71*, 168-181.

Gottholmseder, G., Nowotny, K., Pruckner, G. J., & Theurl, E. (2009). Stress perception and commuting. *Health economics*, *18*(5), 559-576.

Gupta, S., Goren, A., Dong, P., & Liu, D. (2016). Prevalence, awareness, and burden of major depressive disorder in urban China. *Expert review of pharmacoeconomics & outcomes research*, *16*(3), 393-407.

Hansson, E., Mattisson, K., Björk, J., Östergren, P. O., & Jakobsson, K. (2011). Relationship between commuting and health outcomes in a cross-sectional population survey in southern Sweden. *BMC public health*, *11*(1), 1-14.

Hilbrecht, M., Smale, B., & Mock, S. E. (2014). Highway to health? Commute time and well-being among Canadian adults. *World Leisure Journal*, *56*(2), 151-163.

Hong, Xin, JieQuan Li, Fei Xu, Lap Ah Tse, YaQiong Liang, ZhiYong Wang, Ignatius Tak-sun Yu, and Sian Griffiths. 2009. "Physical activity inversely associated with the presence of depression among urban adolescents in regional China." *BMC public health* 9 (1):148.

Huang, Y., Wang, Y. U., Wang, H., Liu, Z., Yu, X., Yan, J., ... & Wu, Y. (2019). Prevalence of mental disorders in China: a cross-sectional epidemiological study. *The Lancet Psychiatry*, *6*(3), 211-224.

Humphreys, D. K., A. Goodman, and D. Ogilvie. 2013. "Associations between active commuting and physical and mental wellbeing." *Preventive Medicine* 57 (2):135-139.

Kessler, R. C., Merikangas, K. R., & Wang, P. S. (2007). Prevalence, comorbidity, and service utilization for mood disorders in the United States at the beginning of the twenty-first century. *Annu. Rev. Clin. Psychol.*, *3*, 137-158.

King, Gary, Robert O Keohane, and Sidney Verba. 1994. *Designing social inquiry: Scientific inference in qualitative research*: Princeton university press.

Kuwahara, K., T. Honda, T. Nakagawa, S. Yamamoto, S. Akter, T. Hayashi, and T. Mizoue. 2015. "Associations of leisure-time, occupational, and commuting physical activity with risk of depressive symptoms among Japanese workers: a cohort study." *Int J Behav Nutr Phys Act* 12:119.

Liu, J., Ettema, D., & Helbich, M. (2022). Systematic review of the association between commuting, subjective wellbeing and mental health. *Travel Behaviour and Society*, *28*, 59-74.

Liu, X., Yan, X., Wang, W., Titheridge, H., Wang, R., & Liu, Y. (2021). Characterizing the polycentric spatial structure of Beijing Metropolitan Region using carpooling big data. *Cities*, *109*, 103040.

Legrain, A., N. Eluru, and A. M. El-Geneidy. 2015. "Am stressed, must travel: The relationship between mode choice and commuting stress." *Transportation Research Part F-Traffic Psychology and Behaviour* 34:141-151.

Ma, Liang, and Runing Ye. 2019. "Does daily commuting behavior matter to employee productivity?" *Journal of Transport Geography* 76:130-141.

Ma, L., Ye, R., & Wang, H. 2021. Exploring the causal effects of bicycling for transportation on mental health. *Transportation research part D: transport and environment*, *93*, 102773.

Mao, Zidan, Dick Ettema, and Martin Dijst. 2016. "Commuting trip satisfaction in Beijing: Exploring the influence of multimodal behavior and modal flexibility." *Transportation Research Part A: Policy and Practice* 94:592-603.

Marmot, Michael. 2005. "Social determinants of health inequalities." *The lancet* 365 (9464):1099-1104.

Martin, A., Y. Goryakin, and M. Suhrcke. 2014. "Does active commuting improve psychological wellbeing? Longitudinal evidence from eighteen waves of the British Household Panel Survey." *Preventive Medicine* 69:296-303.

Milner, A., Badland, H., Kavanagh, A., & LaMontagne, A. D. (2017). Time spent Commuting to work and mental health: evidence from 13 waves of an Australian cohort study. *American journal of epidemiology*, *186*(6), 659-667.

Morris, E. A., & Guerra, E. (2015). Are we there yet? Trip duration and mood during travel. *Transportation research part F: traffic psychology and behaviour*, *33*, 38-47.

Morris, Eric A, and Ying Zhou. 2018. "Are long commutes short on benefits? Commute duration and various manifestations of well-being." *Travel Behaviour and Society* 11:101-110.

National Bureau of Statistics of China. 2016. National 1% Population Sample Survey 2015.

Novaco, Raymond W, Daniel Stokols, Joan Campbell, and Jeannette Stokols. 1979. "Transportation, stress, and community psychology." *American Journal of Community Psychology* 7 (4):361-380.

Novaco, Raymond W, Daniel Stokols, and Louis Milanesi. 1990. "Objective and subjective dimensions of travel impedance as determinants of commuting stress." *American journal of community psychology* 18 (2):231-257.

Olsson, L. E., Gärling, T., Ettema, D., Friman, M., & Fujii, S. (2013). Happiness and satisfaction with work commute. *Social indicators research*, *111*(1), 255-263.

Pearlin, Leonard I, and Alex Bierman. 2013. "Current issues and future directions in research into the stress process." In *Handbook of the sociology of mental health*, 325-340. Springer.

Pearlin, Leonard I, Elizabeth G Menaghan, Morton A Lieberman, and Joseph T Mullan. 1981. "The stress process." *Journal of Health and Social behavior*:337-356.

Pfeiffer, Deirdre, Meagan M Ehlenz, Riley Andrade, Scott Cloutier, and Kelli L Larson. 2020. "Do Neighborhood Walkability, Transit, and Parks Relate to Residents' Life Satisfaction? Insights From Phoenix." *Journal of the American Planning Association* 86 (2):171-187.

Ruger, H., Pfaff, S., Weishaar, H., & Wiernik, B. M. (2017). Does perceived stress mediate the relationship between commuting and health-related quality of life?. *Transportation research part F: traffic psychology and behaviour*, *50*, 100-108.

Shi, Q., & Cao, G. (2020). Urban spillover or rural industrialisation: Which drives the growth of Beijing Metropolitan area. *Cities*, *105*, 102354.

St-Louis, E., Manaugh, K., van Lierop, D., & El-Geneidy, A. (2014). The happy commuter: A comparison of commuter satisfaction across modes. *Transportation research part F: traffic psychology and behaviour*, *26*, 160-170.

Substance Abuse and Mental Health Services Administration. 2012. Results from the 2012 National Survey on Drug Use and Health: Summary of National Findings Washington D.C.: U.S. Department of Health and Human Services.

Sun, B., Lin, J., & Yin, C. (2021). How does commute duration affect subjective well-being? A case study of Chinese cities. *Transportation*, *48*(2), 885-908.

Turner, R. J., Wheaton, B., & Lloyd, D. A. (1995). The epidemiology of social stress. *American Sociological Review*, 104-125.

Voulgaris, Carole Turley, Michael J Smart, and Brian D Taylor. 2019. "Tired of commuting? Relationships among journeys to school, sleep, and exercise among American teenagers." *Journal of Planning Education and Research* 39 (2):142-154.

Wang, K., & Wang, X. (2021). Generational differences in automobility: Comparing America's Millennials and Gen Xers using gradient boosting decision trees. *Cities*, *114*, 103204.

Wang, S., Kamerade, D., Burchell, B., Coutts, A., & Balderson, S. U. (2021). What matters more for employees' mental health: job quality or job quantity?. *Cambridge Journal of Economics*.

Wang, X. (2019). Has the relationship between urban and suburban automobile travel changed across generations? Comparing Millennials and Generation Xers in the United States. *Transportation Research Part A: Policy and Practice*, *129*, 107-122.

Wang, X., Rodríguez, D. A., Sarmiento, O. L., & Guaje, O. (2019). Commute patterns and depression: Evidence from eleven Latin American cities. *Journal of transport & health*, *14*, 100607.

Wener, R., Evans, G. W., & Boately, P. (2005). Commuting stress: Psychophysiological effects of a trip and spillover into the workplace. *Transportation Research Record*, *1924*(1), 112-117.

Wener, Richard E, and Gary W Evans. (2011). "Comparing stress of car and train commuters." *Transportation Research Part F-Traffic Psychology and Behaviour* 14 (2):111-116.

Xiao, C., Yang, Y., & Chi, G. (2020). Does the mental health of migrant workers suffer from long commute time? Evidence from China. *Journal of Transport & Health*, *19*, 100932.

Xu, J., Wang, J., Wimo, A., & Qiu, C. (2016). The economic burden of mental disorders in China, 2005–2013: implications for health policy. *Bmc Psychiatry, 16*(1), 1-9.

Yang, J., Siri, J. G., Remais, J. V., Cheng, Q., Zhang, H., Chan, K. K., . . . Li, X. (2018). The Tsinghua–Lancet Commission on Healthy Cities in China: unlocking the power of cities for a healthy China. *The Lancet, 391*(10135), 2140-2184.

Yin, C., Shao, C., Dong, C., & Wang, X. (2019). Happiness in urbanizing China: The role of commuting and multi-scale built environment across urban regions. *Transportation Research Part D: Transport and Environment*, *74*, 306-317.

Zhu, J., & Fan, Y. (2018). Daily travel behavior and emotional well-being: Effects of trip mode, duration, purpose, and companionship. *Transportation Research Part A: Policy and Practice*, *118*, 360-373.

Zhu, Zhenjun, Zhigang Li, Hongsheng Chen, Ye Liu, and Jun Zeng. 2019. "Subjective well-being in China: how much does commuting matter?" *Transportation* 46 (4):1505-1524.